\documentclass{ws-procs9x6}

\newcommand{\nn}{\nonumber}
\newcommand{\be}{\begin{equation}}
\newcommand{\ee}{\end{equation}}
\newcommand{\ba}{\begin{eqnarray}}
\newcommand{\ea}{\end{eqnarray}}

\def\gev{~{\rm GeV}}

\def\als{\alpha_{\rm s}}

\newcommand{\gsim}{\raisebox{-4pt}{$\,\stackrel{\textstyle
                                                         >}{\sim}\,$}}

\newcommand{\ov}[1]{\overline#1} 

\newcommand{\req}[1]{(\ref{#1})}

\def\xb{\bar{x}}
\def\sh{{s}}
\def\uh{{u}}
\begin{document}
\title{A modern view of wide-angle exclusive scattering}

\author{P.\ Kroll}

\address{Fachbereich Physik, Universit\"at Wuppertal,\\ 
D-42097 Wuppertal, Germany\\
Email: kroll@physik.uni-wuppertal.de}

\maketitle

\abstracts{
The basic theoretical ideas of the handbag mechanism for
wide-angle exclusive scattering reactions are discussed and, with
regard to the present experimental program carried out at JLab, 
its application to Compton scattering is reviewed in some
detail. Results for other wide-angle reactions such as two-photon
annihilations into pairs of hadrons or virtual Compton scattering are
presented as well.\\[0.6em]
Invited talk presented at the N$^*$2002 workshop on the physics of excited
nucleons, Pittsburgh, October 2002 } 
%%%%%%%%%%%%%%%%%%%%%%%%%%%%%%%%%%%%%%%%%%%%%%%%%%%%%%%%%%%%%%%%%%%%%%%%%
\section{Introduction}
\begin{figure}[t]
\begin{center}
\includegraphics[width=4.3cm,bbllx=45pt,bblly=230pt,bburx=545pt,
bbury=615pt,clip=true]{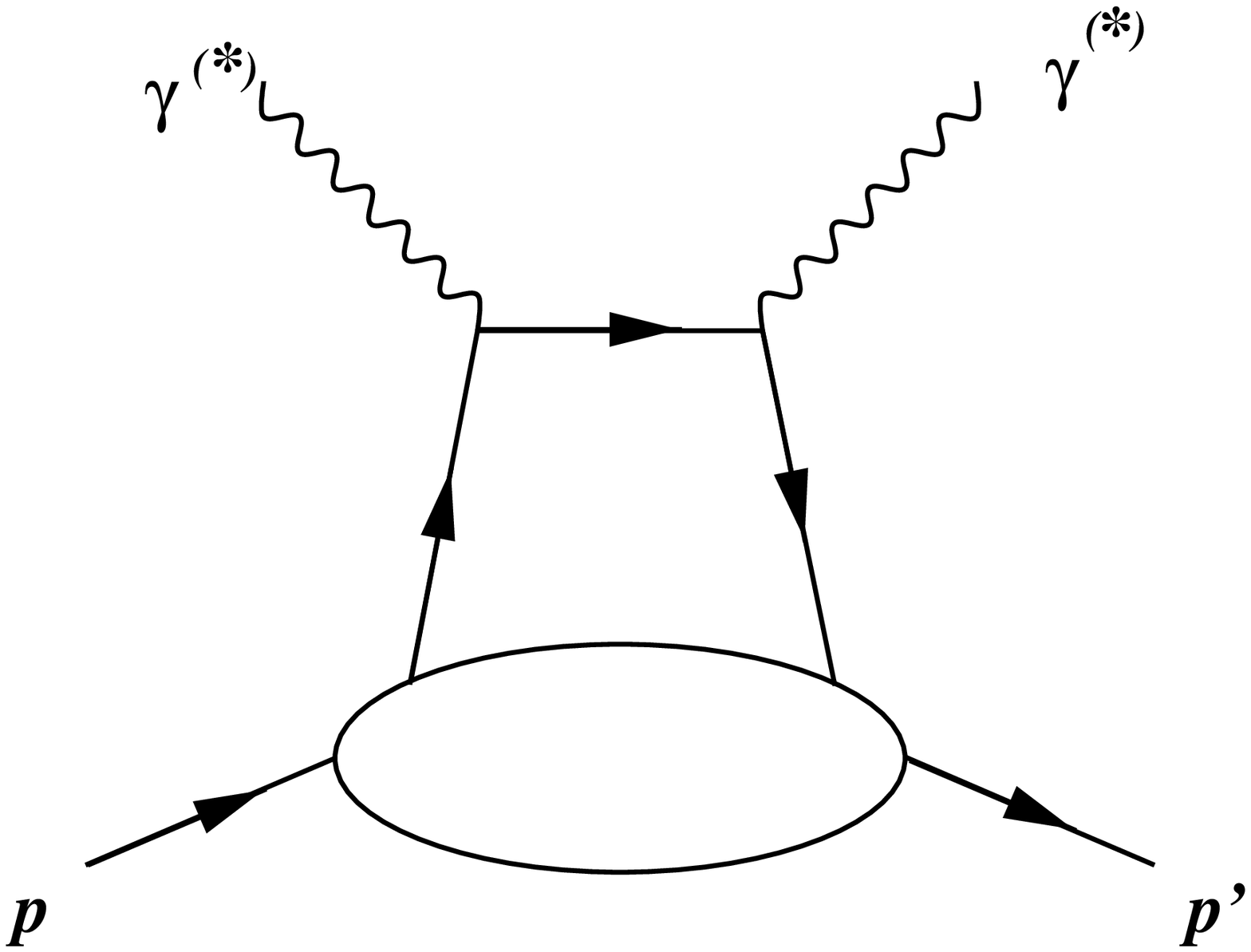} 
%\includegraphics[width=4.5cm,bbllx=73pt,bblly=525pt,bburx=510pt,
%bbury=795pt,clip=true]{Fig-gpd/comp-pqcd.ps} 
\includegraphics[width=4.3cm,bbllx=182pt,bblly=410pt,bburx=400pt, 
bbury=610pt,clip=true]{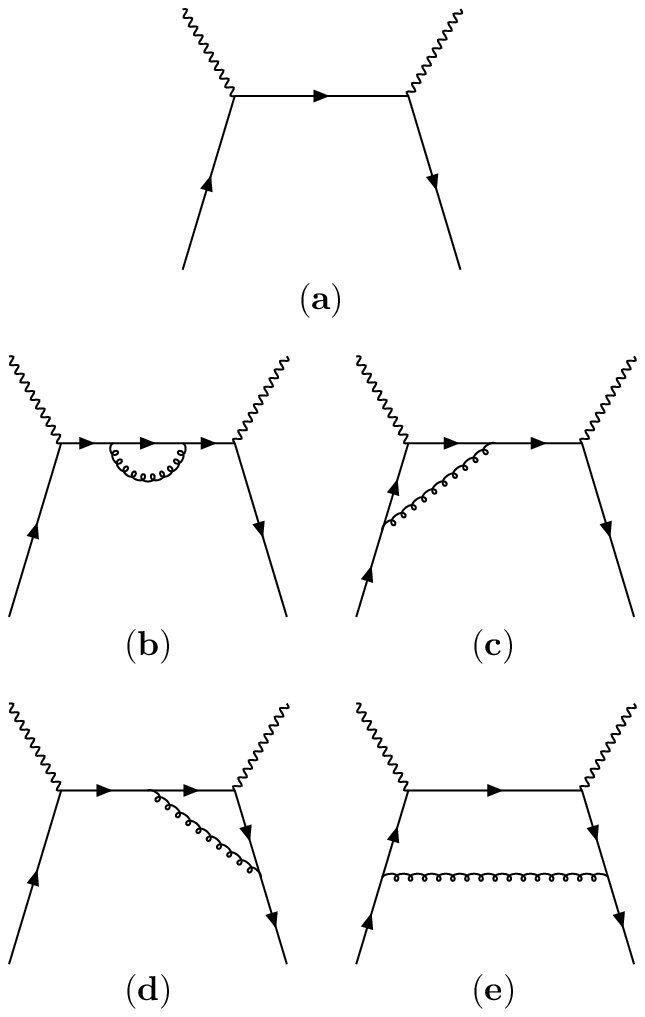} 
\caption{{}Handbag diagram for Compton scattering (left) and sample 
NLO pQCD Feynman graphs for its parton subprocess 
$\gamma q\to \gamma q$.} 
\label{fig:handbag}
\end{center}
\end{figure}

%%%%%%%%%%%%%%%%%%%%%%%%%%%%%%%%%%%%%%%%%%%%%%%%%%%%%%%%%%%%%%%%%%%%%%%%%
Recently a new approach to hard Compton scattering (CS) off protons
has been proposed where the process amplitudes factorize into a hard
parton-level subprocess, Compton scattering off quarks, and generalized
parton distributions (GPDs) which encode the soft physics (see Fig.\
\ref{fig:handbag}). This so-called handbag mechanism applies to deep
virtual Compton scattering (DVCS) \cite{rad97} characterized by a large
virtuality, $Q^2$, of the incoming photon and a small squared
invariant momentum transfer, $-t$, from the incoming to the outgoing
proton ($-t/Q^2\ll 1$). Subsequently it has been realized that the
handbag mechanism also applies to wide-angle CS
\cite{rad98,DFJK1} for which $-t$ (and $-u$) are large but the photon
virtuality is small or even zero ($-Q^2/t\ll 1$). It is believed now 
that the handbag mechanism is the relevant physics for a large number
of deep virtual and wide-angle exclusive reactions such as 
electroproduction of mesons or two-photon annihilations into pairs of
hadrons. 

Wide-angle exclusive reactions and in particular real CS are the
subject of my talk. The handbag mechanism in CS is described in
Sect.\ 2. The large $-t$ behaviour of GPDs and form factors is
discussed in Sect.\ 3 and predictions for CS are given. Characteristic
results for other wide-angle exclusive processes are presented in
Sect.\ 4.    

%%%%%%%%%%%%%%%%%%%%%%%%%%%%%%%%%%%%%%%%%%%%%%%%%%%%%%%%%%%%%%%%%%%%%%%%%
\section{Wide-angle Compton scattering}
%%%%%%%%%%%%%%%%%%%%%%%%%%%%%%%%%%%%%%%%%%%%%%%%%%%%%%%%%%%%%%%%%%%%%%%%%
For Mandelstam variables $s$, $-t$ and $-u$ that are large as compared
to a typical hadronic scale $\Lambda^2$ where $\Lambda$ being of order
$1\, \gev$, it can be shown that the handbag diagram shown in
Fig.\ \ref{fig:handbag} describes CS. To see this it is of advantage 
to work in a symmetrical frame which is a c.m.s rotated in such a way
that the momenta of the incoming ($p$) and outgoing ($p'$) proton
momenta have the same light-cone plus components. In this frame the
skewness, defined as 
\be 
\xi = \frac{(p - p')^+}{(p + p')^+}\,,
\ee
is zero. The bubble in the handbag is viewed as a sum over all possible 
parton configurations as in deep ineleastic lepton-proton scattering (DIS). 
The crucial assumptions in the handbag approach are that of restricted
parton virtualities, $k_i^2<\Lambda^2$, and of intrinsic transverse
parton momenta, ${\bf k_{\perp i}}$, defined with respect to their
parent hadron's momentum, which satisfy $k_{\perp i}^2/x_i
<\Lambda^2$, where $x_i$ is the momentum  fraction parton $i$ carries.   
 
One can then show \cite{DFJK1} that the subprocess Mandelstam variables
$\hat{s}$ and $\hat{u}$ are the same as the ones for the full process,
Compton scattering off protons, up to corrections of order
$\Lambda^2/t$:
\ba
\hat{s}=(k_j+q)^2 \simeq (p+q)^2 =s\,, \quad 
\hat{u}=(k_j-q')^2 \simeq (p-q')^2 =u\,.
\ea
The active partons, i.e.\ the ones to which the photons couple, are
approximately on-shell, move collinear with their parent hadrons and
carry a momentum fraction close to unity, $x_j, x_j' \simeq 1$.
Thus, like in DVCS, the physical situation is that of a hard
parton-level subprocess, $\gamma q\to \gamma q$, and a soft emission
and reabsorption of quarks from the proton. The light-cone helicity 
amplitudes \cite{diehl01} for wide-angle CS then read
\ba
{M}_{\mu'+,\,\mu +}(s,t) &=& \; \frac{e}{2}
     \left[\, { T}_{\mu'+,\,\mu+}(\sh,t)\,(R_V(t) + R_A(t))\,\right.\nn\\[0.5em]
&&\qquad\left.  + \,\,  { T}_{\mu'-,\,\mu-}(\sh,t)\,(R_V(t) - R_A(t)) \right]  
                                                 \,, \label{ampl}\\[0.5em]
 { M}_{\mu'-,\,\mu +}(s,t) &=& \;- \frac{e}{2} \frac{\sqrt{-t}}{2m} 
         \left[\,  T_{\mu'+,\,\mu+}(\sh,t)\, 
         + \,  { T}_{\mu'-,\,\mu-}(\sh,t)\, \right] \,R_T(t)\,.\nn
\ea
$\mu,\, \mu'$ denote the helicities of the incoming and outgoing
photons, respectively. The helicities of the protons in $ { M}$ and
of the quarks in the hard scattering amplitude $ T$ are labeled by their
signs. $m$ denotes the mass of the proton. The hard scattering has
been calculated to next-to-leading order perturbative QCD \cite{HKM}, 
see Fig.\ \ref{fig:handbag}. To this order the gluonic subprocess, 
$\gamma g\to \gamma g$, has to be taken into account as well. The 
form factors $R_i$ represent $1/\xb$-moments of GPDs at zero skewness. 
This representation which requires the dominance of the plus 
components of the proton matrix elements, is a non-trivial feature
given that, in contrast to DIS and DVCS, not only the plus components
of the proton momenta but also their minus and transverse components
are large here. It is interesting to note that the DVCS 
amplitudes~\cite{rad97}
\ba
 M^{\rm DVCS} &\propto& \bar{u}(p')\gamma^+ u(p) 
               \int_{-1}^1 d\xb H(\xb,\xi,t) \Big[ \frac1{\xb-\xi
     +i\epsilon} + \frac1{\xb+\xi-i\epsilon} \Big]\nn\\[0.2em]
  && \;+\quad  E\,, \widetilde{H}\,, \widetilde{E}\, - {\rm terms}\,,
\label{dvcs}
\ea
although being derived for large $Q^2$ and small $-t$, embodies the
wide-angle amplitudes \req{ampl} as can easily been seen by setting
$\xi=0$ and evaluating the kinematical factors in front of the
integral at large $-t$. The integrals over $H$, $E$ and $\widetilde{H}$ 
turn into the form factors $R_V$, $R_T$ and $R_A$, respectively. The
GPD $\widetilde{E}$ does not contribute at $\xi=0$.  

The handbag amplitudes \req{ampl} lead to the following result for the
Compton cross section  
\ba
\frac{d\sigma}{dt} &=& \frac{d\hat{\sigma}}{dt} \left\{ \frac12\, \big[
R_V^2(t)\,(1+\kappa_T^2) + R_A^2(t)\big] \right.\nn\\
&&\left. \quad  - \frac{\uh\sh}{\sh^2+\uh^2}\, \big[R_V^2(t)\,(1+\kappa_T^2) 
                - R_A^2(t)\big]\right\} + O(\alpha_s)\,,
\label{dsdt}
\ea
where $d\hat{\sigma}/dt$ is the Klein-Nishina cross section for
CS of massless, point-like spin-1/2 particles of
charge unity. The parameter $\kappa_T$ is defined as 
\be
\kappa_T= \frac{\sqrt{-t}}{2m}\, \frac{R_T}{R_V} 
\ee
Another interesting observable in CS is the helicity
correlation, $A_{LL}$,  between the initial state photon and proton
or, equivalently, the helicity transfer, $K_{LL}$, from the incoming
photon to the outgoing proton. In the handbag approach one obtains
\cite{HKM,DFJK2} 
\be
A_{LL}=K_{LL}\simeq \frac{\sh^2 - \uh^2}{\sh^2 + \uh^2}\, 
                    \frac{R_A}{R_V} + O(\kappa_T,\alpha_s)\,,
\label{all}
\ee  
where the factor in front of the form factors is the corresponding
observable for $\gamma q\to \gamma q$. The result \req{all} is a
robust prediction of the handbag mechanism, the magnitude of the
subprocess helicity correlation is only diluted  somewhat by the ratio
of the form factors $R_A$ and $R_V$. On the other hand, the helicity 
correlation for sideways polarized protons (i.e.\ perpendicular to the
proton's three-momentum and in the scattering plane), is very
sensitive to details of the approach. It reads 
\be
A_{LS}=-K_{LS}\simeq \frac{-t}{\sh-\uh}\, \frac{R_A}{R_V}\,
  \kappa_T\, \left[1+ \frac{2m}{\sqrt{\sh}}\,
      \frac{\sqrt{-t}}{\sqrt{\sh}+\sqrt{\uh}} \, \kappa_T^{-1}\right] +
      O(\als)\,.
\ee
%%%%%%%%%%%%%%%%%%%%%%%%%%%%%%%%%%%%%%%%%%%%%%%%%%%%%%%%%%%%%%%%%%%%%%%%
\section{The large-$t$ behaviour of GPDs}
\label{sect:model}
%%%%%%%%%%%%%%%%%%%%%%%%%%%%%%%%%%%%%%%%%%%%%%%%%%%%%%%%%%%%%%%%%%%%%%%% 
In oder to make actual predictions for CS however models for the 
soft form factors or rather for the underlying GPDs are required.
A first attempt to parameterize the GPDs $H$ and $\widetilde{H}$  
at zero skewness has been given in \cite{rad98,DFJK1,HKM} 
\ba
H^a(\xb,0;t) &=& \exp{\left[a^2 t
        \frac{1-\xb}{2\xb}\right]}\, q_a(\xb)\,,\nn\\ 
\widetilde{H}^a(\xb,0;t) &=& \exp{\left[a^2 t
        \frac{1-\xb}{2\xb}\right]}\, \Delta q_a(\xb)\,,
\label{gpd}
\ea
where $q(\xb)$ and $\Delta q(\xb)$ are the usual unpolarized and polarized
parton distributions in the proton~\footnote{The parameterization
\req{gpd} can be motivated by overlaps of light-cone wave functions
which have a Gaussian $\vec{k}_\perp$ dependence
\cite{rad98,DFJK1,DFJK3}.}. $a$, the transverse size of the 
proton, is the only free parameter and even it is restricted to the
range of about 0.8 to 1.2 $\gev^{-1}$. Note that $a$ essentially refers to 
the lowest Fock states of the proton which, as phenomenological
experience tells us, are rather compact. The model (\ref{gpd}) is
designed for large $-t$. Hence, forced by the Gaussian in (\ref{gpd}),
large $\xb$ is implied, too. Despite of this the normalization of 
the model GPDs at $t=0$ is correct. 

\begin{figure}[t]
\begin{minipage}{0.46\textwidth}
\begin{center} 
\includegraphics[width=3.0cm,bbllx=90pt,bblly=30pt,bburx=590pt,
bbury=635pt,angle=-90,clip=true]{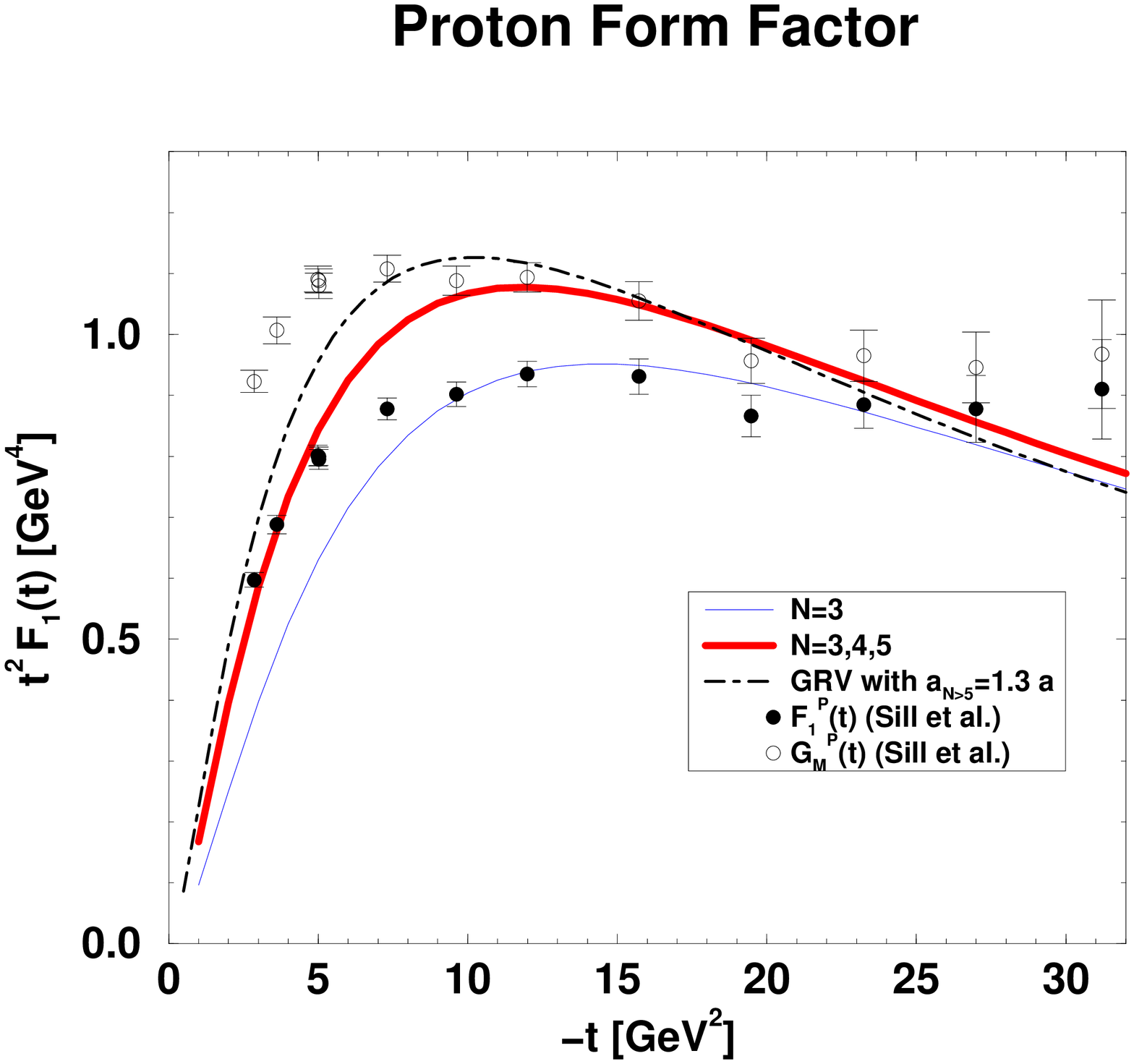}%\hspace*{1cm}
\end{center}
\end{minipage}
\begin{minipage}{0.46\textwidth}
\begin{center}
\includegraphics[width=4.5cm,bbllx=27pt,bblly=47pt,bburx=398pt, 
bbury=295pt,clip=true]{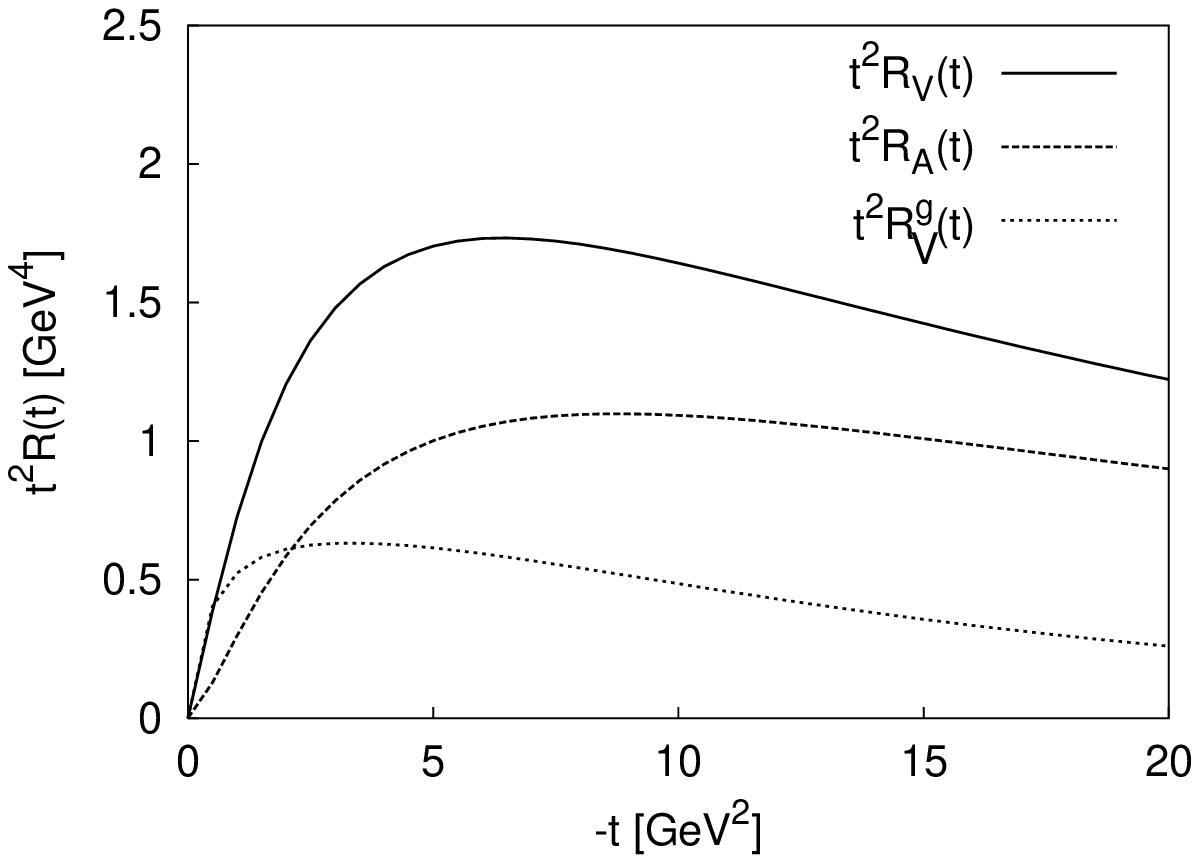}
\end{center}
\end{minipage} 
\caption{The Dirac form factor of the proton (left) and the Compton form
factors (right). The figures are taken from Refs.~\protect\cite{DFJK1,HKM}, data
are taken from Ref.\ \protect\cite{sill}.}
\label{fig:form}
\end{figure}

With the model GPDs \req{gpd} at hand one can evaluate the various form
factors by taking appropriate moments, e.g.\
\be
F_1=\sum_q e_q \int_{-1}^1 d\xb H^q(\xb,0;t)\,,\quad
R_V=\sum_q e_q^2 \int_{-1}^1 \frac{d\xb}{\xb} H^q(\xb,0;t)\,.
\label{formfactors}
\ee
Results for the form factors are shown in Fig.\ \ref{fig:form}. Obviously,  
as the comparison with experiment \cite{sill} reveals, the model GPDs 
work quite well in the case of the Dirac form factor. The 
scaled form factors $t^2 F_1$ and $t^2 R_i$ exhibit broad maxima which
mimick dimensional counting in a range of $-t$ from, say, $5$ to about
$20\,\gev^2$. The position of the maximum of any of the scaled form
factors is approximately located at \cite{DFJK2} 
\be
t_0 \simeq -4 a^{-2}\, \left\langle
                          \frac{1-\xb}{\xb}\right\rangle^{-1}_{F(R)}\,.
\label{max-pos}
\ee\
The mildly $t$-dependent mean value $\langle (1-\xb)/\xb\rangle$ comes
out around $1/2$. A change of $a$ moves the position of the maximum of 
the scaled form factors but leaves their magnitudes essentially unchanged. 
 
The Pauli form factor $F_2$ and its Compton analogue $R_T$ contribute
to proton helicity flip matrix elements and are related to the GPD $E$
analogously to \req{formfactors}. This connection suggests that, at
least for not too small values of $-t$, $R_T/R_V$ roughly behaves as
$F_2/F_1$. Thus, on the basis of the SLAC data~\cite{slac} on 
$F_2/F_1$, one expects $R_T/R_V\propto m^2/t$ while the recent JLab
data \cite{gayou} rather indicate a behaviour as $\propto m/\sqrt{-t}$. 
If the first estimate is correct the contribution from the form factor
$R_T$ to Compton scattering can be ignored while in the second case it 
is to be taken into account since it contributes to the same order in
$\Lambda/\sqrt{-t}$ as the other form factors. Since it is not yet
clear which behaviour is the correct one, predictions for Compton 
observables are given for two different scenarios. Both $R_T$ and
$\als$ corrections are omitted in scenario B but taken into account in 
A where the ratio $\kappa_T$ is assumed to have a value of 0.37 as
estimated from the JLab form factor data \cite{gayou}.

Employing these model GPDs and the corresponding form factors, various
Compton observables can be calculated \cite{DFJK1,HKM,DFJK2}. The
predictions for the differential cross section are in fair agreement
with experiment. The approximative $s^6$-scaling of the predictions is
related to the broad maximum the scaled form factors exhibit, see
Fig.\ \ref{fig:form}. JLab will provide accurate cross section data
soon which will allow a detailed examination of the handbag mechanism.  
\begin{figure}[t]
\begin{center}
\includegraphics[width=4.7cm,bbllx=33pt,bblly=47pt,bburx=400pt, 
bbury=295pt,clip=true]{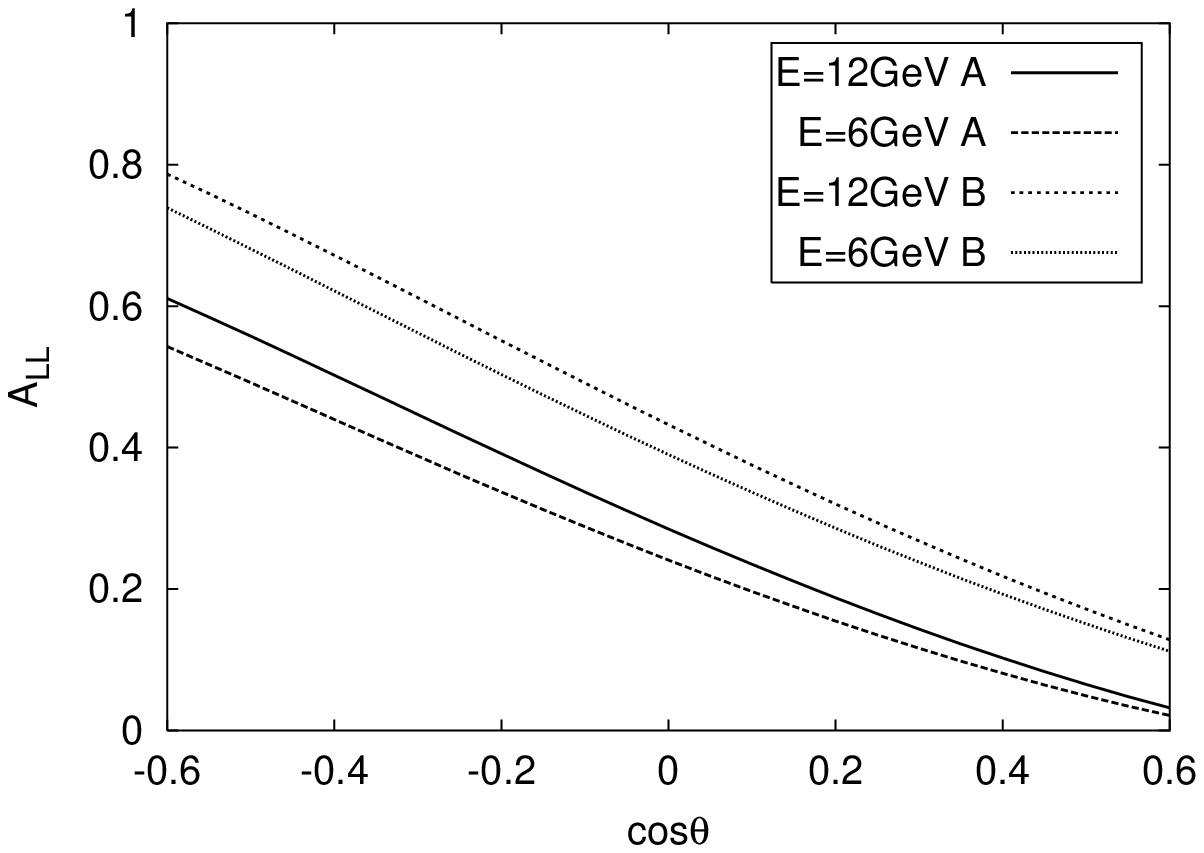}
\includegraphics[width=4.2cm,bbllx=322pt,bblly=55pt,bburx=565pt, 
bbury=246pt,clip=true]{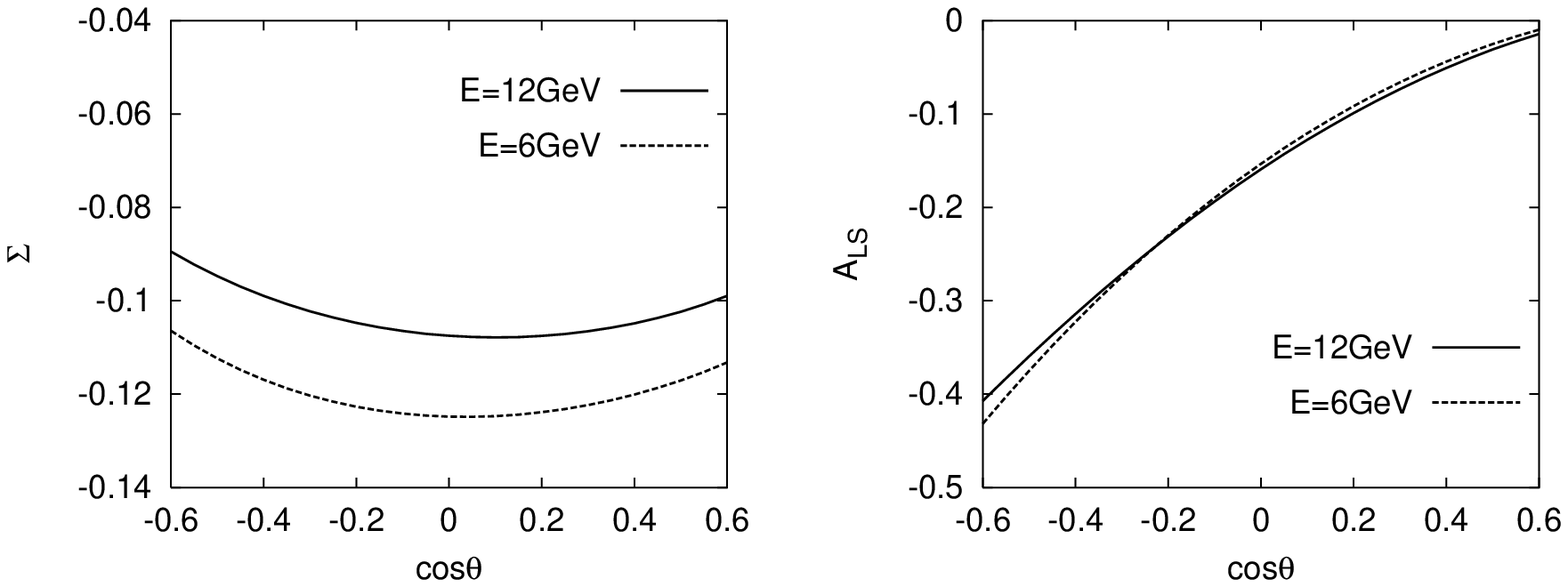} 
\caption{Predictions for the helicity correlations $A_{LL}=K_{LL}$
(left) and $A_{LS}=-K_{LS}$ (right)\protect\cite{HKM}. NLO
corrections and the tensor form factor are taken into account 
(scenario A), in scenario B they are neglected.} 
\label{fig:cross}
\end{center}
\end{figure}
Predictions for $K_{LL}$ and $K_{LS}$ are shown in Fig.\ \ref{fig:cross}.
The JLab E99-114 collaboration \cite{nathan} has presented a first
measurement of $K_{LL}$ and $K_{LS}$ at a c.m.s.\ scattering angle of
$120^\circ$ and a photon energy of $3.23 \gev$. These still
preliminary data points are in fair agreement with the predictions
from the handbag given the small energy at which they are
available. The kinematical requirement of the handbag mechanism 
$s,\; -t,\; -u \gg \Lambda^2$ is not well satisfied and therefore one
has to be aware of large dynamical and kinematical corrections. Among
them there are proton mass effects which have been investigated in
Ref.\ \cite{DFHK}. 

There is an alternative to the handbag factorization. This is
the leading-twist scheme \cite{bro80} where all valence quarks  
the involved hadrons are made up participate in the hard scattering
and not just a single one. Although it is believed for good reasons 
that the leading-twist scheme dominates for asymptotically large $s$, 
it is not clear theoretically which of the two approaches provides the 
appropriate description of wide-angle Compton scattering at, say, 
$-t \simeq 10 \gev^2$. The ultimate decision is to be made by
experiment. In fact, leading-twist calculations, e.g.\ \cite{dixon}, 
reveal difficulties with the size of the Compton cross section, the 
numerical results are way below experiment. Moreover, the
leading-twist approach leads to a negative value for $K_{LL}$ at
angles larger than $90^\circ$ where the handbag predictions are
positive (see Fig.\ \ref{fig:cross}). $K_{LS}$ is zero to
leading-twist order. 
    
For large $-t$ and $\xb\gsim 0.6$, the zero-skewness GPDs \req{gpd}
can be parameterized as 
\be
H^q(\xb,0,t)\simeq f_q\, \xb\; (1-\xb)^{\,b_q}\,   
                \exp{\big[\frac{a^2 t}{2}\, \frac{1-\xb}{\xb}\big]}   \,,
\label{app}
\ee
and analogously for the other ones; evolution is ignored for simplicity.
The various form factors, see for instance \req{formfactors}, imply
integrals over GPDs from 0 to 1. However, for $-t \gsim 10 \gev^2$, the
exponential in \req{app} cuts off the small $\xb$ region and, as can 
easily be checked numerically, it suffices to integrate from 0.6 to 1. 
Hence, one can work out the large $-t$ behaviour of the form factors 
from the parameterization \req{app}. 

For very large values of $-t$, well above 
$100\,\gev^2$, the form factors behave as \cite{DFJK1,BK}
\be
F_1^q\,,\, R_{V,A}^q \propto (-1/t)^{\,b_q+1}\,.
\label{asy}
\ee 
This correspondence between the large-$\xb$ behaviour of the parton
distributions and the large $-t$ behaviour of the form factors is
analogous to the Drell-Yan-West relation \cite{DYW}. The asymptotic behaviour
\req{asy} emerges very slowly; for $-t$ near $10 \gev^2$ the form
factors effectively behave $\propto t^2$ as can be seen from Fig.\
\ref{fig:form}.

Using $b_u=b_d\simeq 3$ and $b_{sea}\simeq 7$ in agreement with overlaps
evaluated from SU(6) symmetric wave functions \cite{DFJK1,BK}
\footnote{In the phenomenological parton distributions, see e.g.\ 
Ref.~\cite{GRV}, $b_d$ is rather 4 than 3 with the 
consequence of suppressed $d$-quark contributions to the
electromagnetic proton and neutron form factors for large $-t$. In 
the large-$\xb$ region however, the errors in the phenomenological parton
distributions are substantial.}, one
sees that the active quarks in the handbag are valence quarks for large $-t$.
This does not imply that the proton is only made of
valence quarks; the bubble in the handbag, see Fig.\ \ref{fig:handbag}, 
represents a sum over all parton configurations allowed by the
conservation laws. Evaluating the form factors from \req{app} and the 
above powers $b_q$, one finds for the ratios of neutron over proton
form factors  
\be    
\frac{d}{u} \to \rho=\frac{f_d}{f_u}\,, \qquad
\frac{F_{1n}}{F_{1p}} \to \frac{\rho + e_d/e_u}{1 + e_d/e_u\, \rho}\,, \qquad
\frac{R_{1n}}{R_{1p}} \to \frac{\rho + (e_d/e_u)^2}{1 + (e_d/e_u)^2\, \rho}\,, 
\label{neutron}
\ee
which approximately hold for  $-t\gsim 10 \gev^2$. Since the vector
form factor dominates the Compton cross section, the ratio of neutron
over proton cross sections is approximatively given by
\be
d\sigma_n/d\sigma_p \to \Big[\frac{\rho + (e_d/e_u)^2}{1 +
    (e_d/e_u)^2\, \rho}\Big]^2\,.
\label{cross-neutron}
\ee
Measurements of the parton distributions for $\xb\gsim 0.6$ and of the
neutron form factors at sufficiently large $-t$ would thus allow
further tests of the handbag mechanism. It is to be stressed that the
relations \req{neutron} do not rely on details of the
parameterization \req{app}, required is only a sufficiently strong
suppression of contributions from the low-$\xb$ region. The asymptotic
behaviour \req{asy}, however, demands more. The effective range of 
$\xb$ must shrink to unity with increasing $-t$ (\,$[1+c/t,1]$ with $c>0$\,).
%%%%%%%%%%%%%%%%%%%%%%%%%%%%%%%%%%%%%%%%%%%%%%%%%%%%%%%%%%%%%%%%%%%%%%%%
\section{The handbag mechanism in other wide-angle reactions}
%%%%%%%%%%%%%%%%%%%%%%%%%%%%%%%%%%%%%%%%%%%%%%%%%%%%%%%%%%%%%%%%%%%%%%%%
The handbag approach has been applied to several other high-energy
wide-angle reactions. Thus, as shown in Ref.\ \cite{DFJK2}, the
calculation of {\it real Compton scattering} can be straightforwardly  
extended to {\it virtual Compton scattering} provided $Q^2/-t \ll
1$. 
The handbag approach also applies to reactions like $\gamma p \to
\gamma \Delta (N^*)$. New GPDs, parameterizing the soft proton-$\Delta (N^*)$ 
matrix elements, occur in these reactions \cite{Frankfurt}. For the 
wide-angle region, however, such processes have not yet been calculated.  

{\it Photo- and electroproduction of mesons} have also been discussed
within the handbag approach~\cite{hanwen} using, as in deep virtual
electroproduction~\cite{dvem}, a one-gluon exchange mechanism for the
generation of the meson. The normalization of the photoproduction
cross section is not yet understood. Either vector meson dominance
contributions are still predominant or the generation of the meson
by the exchange of a hard gluon underestimates the handbag
contribution. Despite of this the handbag contribution to photo- and 
electroproduction has several interesting properties which perhaps 
survive an improvement of the approach.  For instance, the helicity 
correlation $\hat{A}_{LL}$ for the subprocess $\gamma q \to \pi q$ is
the same as for $\gamma q \to \gamma q$, see (\ref{all}). $A_{LL}$ 
for the full process is diluted by form factors similar to the case of
Compton scattering. Another interesting result is 
the ratio of the cross sections for the photoproduction of $\pi^+$ and 
$\pi^-$ which is approximately given by
\be
\frac{d\sigma(\gamma n\to \pi^- p)}{d\sigma(\gamma p\to \pi^+ n)} \simeq
\left[\frac{e_d \uh + e_u \sh}{e_u \uh + e_d \sh}\right]^2\,.
\label{pi-ratio}
\ee
The form factors which, for a given flavor, are the same as those
appearing in CS, cancel in the ratio. The prediction~\req{pi-ratio} is
in fair agreement with a recent JLab measurement \cite{zhu} 
which, at $90^\circ$, provides values of $1.73\pm 0.15$ and $1.70\pm 0.20$
for the ratio at beam energies of $4.158$ and $5.536 \gev$, respectively.

{\it Elastic hadron-hadron scattering} can be treated as well. 
Details have not yet been worked out but it has been shown that form
factors of the type discussed in Sect.\ \ref{sect:model} control
elastic scattering, too~\cite{DFJK2}. The experimentally observed
scaling behaviour of these cross sections can be attributed 
to the broad maxima the scaled form factors show, see Fig.\
\ref{fig:form} and Eq.\ \req{max-pos}. 

{\it Two-photon annihilations into pairs of hadrons} can also be
calculated, the arguments for handbag factorization hold as well as
has recently been shown in Ref.\ \cite{DKV2} (see also Ref.~\cite{weiss}). 
The cross section for the production of a pair of pseudoscalar mesons
or baryons  read 
\ba
\frac{d\sigma}{dt}(\gamma\gamma\to M\ov{M}) &=& 
     \frac{8\pi\alpha^2_{\rm elm}} 
              {s^2 \sin^4 \theta} \big|R_{M\ov{M}}(s)\big|^2\nn\\
\frac{d\sigma}{dt}\,(\,\gamma\gamma\,\to\,\, B\ov{B}\,) &=& 
      \frac{4\pi\alpha^2_{\rm elm}} 
        {s^2 \sin^2 \theta} \Big\{ \big|R_A^B(s)+ R_P^B(s)\big|^2\nn\\
                  &+& \cos^2\theta\,\big|R_V^B(s)\big|^2\,+\, 
                     \frac{s}{4m^2}\,\big| R_P^B(s)\big|^2 
                                       \Big\}\,.
\ea
In analogy to Eq.\ \req{formfactors} the form factors represent
integrated two-hadron distribution amplitudes which are time-like
versions of GPDs. The angle dependencies are in fair agreement with
experiment.

A characterisic feature of the handbag mechanism in the time-like
region is the intermediate $q\ov{q}$ state implying the absence of 
isospin-two components in the final state. A consequence of this property is 
\be
\frac{d\sigma}{dt}(\gamma\gamma\to \pi^0\pi^0) = 
                 \frac{d\sigma}{dt}(\gamma\gamma\to \pi^+\pi^-)\,,
\ee
which is independent of the soft physics input and is, in so
far, a robust prediction of the handbag approach. 
The absence of the isospin-two components combined with flavor
symmetry allows one to calculate the cross sections for other $B\ov{B}$
channels using the form factors for $p\ov{p}$ as the only soft physics
input. 
%%%%%%%%%%%%%%%%%%%%%%%%%%%%%%%%%%%%%%%%%%%%%%%%%%%%%%%%%%%%%%%%%%%%%%%%
\section{Summary}
%%%%%%%%%%%%%%%%%%%%%%%%%%%%%%%%%%%%%%%%%%%%%%%%%%%%%%%%%%%%%%%%%%%%%%%%
I have reviewed the theoretical activities on applications of the
handbag mechanism to wide-angle scattering. There are many interesting
predictions, some are in fair agreement with experiment, others still
awaiting their experimental examination. It seems that the handbag
mechanism plays an important role in wide-angle exclusive reactions
for momentum transfers of the order of $10\gev^2$. However, before we
can draw firm conclusions more experimental tests are needed.
The leading-twist approach, on the other hand, typically provides
cross sections which are way below experiment. As is well-known the
cross section data for many hard exclusive processes exhibit
approximate dimensional counting rule behaviour. Infering
from this fact the dominance of the leading-twist contribution is
premature. The handbag mechanism can explain this approximate
power law behaviour (and often the magnitude of the cross sections),
too. It is attributed to the broad maxima the scaled form factors
show and, hence, reflects the the transverse size of the lowest Fock
states of the involved hadrons.  

I finally emphasize that the structure of the handbag amplitude,
namely its representation as a product of perturbatively calculable
hard scattering amplitudes and $t$-dependent form factors is the
essential result. Refuting the handbag approach necessitates  
experimental evidence against this factorization.

%%%%%%%%%%%%%%%%%%%%%%%%%%%%%%%%%%%%%%%%%%%%%%%%%%%%%%%%%%%%%%%%%%%%%%%%%%%

\end{document}